\documentclass[12pt]{article}

\usepackage{graphicx}
\usepackage{bm}

\begin{document}

\centerline{\Huge\bf Quantum time ordering and degeneracy}
\vskip 1em
\centerline{\LARGE \bf I: Time ordering in quantum mechanics}

\vskip 2em

\centerline{\it J.H. McGuire$^\dagger$, A.L. Godunov$^\ddagger$, Kh.Kh. Shakov$^\dagger$,}
\centerline{\it Kh.Yu. Rakhimov$^{\dagger}$\footnote{Permanent address:
Department of Heat Physics, Uzbekistan Academy of Sciences,
28 Katartal St., Tashkent 700135, Uzbekistan} and A. Chalastaras$^\dagger$}

\vskip 1em

\centerline{$^\dagger$ Physics Department, Tulane University, New Orleans, LA 70118 USA.}
\centerline{$^\ddagger$ Physics Department, Old Dominion University, Norfolk, VA, 23529 USA.}

\date{\today}

\begin{abstract}
We examine how time ordering works in quantum mechanics and in classical mechanics.
\end{abstract}

\section{Introduction}

Causality and entropy both tell us something about how time works.
Cause happens before effect.
In non-dissipative systems physical observables are generally invariant
under the reversal of time.
Irreversible dissipation gives an observable direction
to the flow of time, i.e. entropy provides the `arrow of time'.
In this paper we examine the nature of time ordering of interactions
in non-dissipative quantum systems.  We associate this time ordering with causality.

Classically, the mathematical function that connects cause and effect is called
a Green's function, $G$.
This function describes how dynamic systems evolve from position ($\vec{r}_0,t_1$)
to ($\vec{r},t_2$).
This Green's function is used to describe, for example, the time sequence in
which interactions occur.
Certain conditions are usually imposed on the Greens' function \cite{MF}.
One such condition is {\it causality}.  If an interaction occurs at $t_1$,
no effect of this interaction should be present at an earlier time.
This imposes the condition that $G = dG/dt = 0$ for $t_2 < t_1$.
Conversely, an event or interaction at $t_1$ can affect the system at a later
time, $t_2$.  This is a classical example of time non-locality \cite{Jackson}.
The {\it reciprocity} condition imposes a symmetry between propagation
from ($\vec{r}_0,t_1$) to ($\vec{r},t_2$), and time reversed propagation
from ($\vec{r},t_2$) to ($\vec{r}_0,t_1$).  The condition is that
$G(\vec{r},t_2;\vec{r}_0,t_1)= G(\vec{r}_0,-t_1;\vec{r},-t_2)$.
The effect of an interaction at ($\vec{r},t_2$) from ($\vec{r}_0,t_1$)
is equal to the effect of an interaction at ($\vec{r}_0,-t_1$) from ($\vec{r},-t_2$),
where $-t_2$ now precedes $-t_1$, cleverly satisfying causality.
Moreover to specify a unique physical solution to a differential equation
{\it boundary and/or initial conditions} are required.
In solving Newton's equations for particles, one often
specifies the initial values of $\vec{r}_0$ and $d \vec{r}_0 /dt$
to determine a unique trajectory, $\vec{r}(t)$ for a particle.
For scattering of waves most typical is the boundary condition that the scattered waves
be outgoing, corresponding to an $e^{ikr}/r$ scattered term.
Alternatively, incoming scattering waves could be used\footnote{Actually
doing such a time reversed experiment could be very difficult.}, corresponding to $e^{-ikr}/r$.
If waves reflect from some finite boundary, combinations of
outgoing and incoming waves may be appropriate. One example \cite{MF}
is a standing wave, corresponding to $\cos kr/r$.
One way or another the flow of time must be clearly specified.

Classical waves are usually not well localized.
Waves can continue for a long time, and can be modified.
Radio wave propagation is an example.
The modulation of a high frequency, so-called carrier wave
can be used to transmit information, such as the news.
Wave modulation, based on interference of waves of different frequencies,
is distributed, i.e. non-local.
Classically all precursors of a wave are restricted \cite{Jackson}
to group velocities less than the speed of light\footnote{In anamolous dispersion
a zero group velocity can achieved \cite{Jackson,mair02} when $d\omega/dk = 0$.
This can also be realized in a degenerate quantum system where all energies are
the same so that $dE/dk = 0$.}.
The properties of Green's functions for waves \cite{MF,Jackson} are similar to
those for particles.

Quantum mechanics brings together particles and waves by use of finite wavepackets.
This can also be done classically.
Wavepackets have properties of {\it both} particles and waves.
As a consequence, quantum wavepackets may not be perfectly localized in both time and frequency.
A particle-like wavepacket requires many frequencies so the waves
can cancel except at small distances.
And a wave-like wavepacket oscillating with nearly a single frequency (small $\Delta f$),
must be spread over time (large $\Delta t$).
Here $c \Delta t$ is the size of a wavepacket moving at a speed $c$.
This illustrates the band-width theorem, $\Delta f \Delta t \geq 1/4 \pi$.
Unlike in classical mechanics, in quantum mechanics the energy is related to
the frequency, e.g. $E = hf$.
In quantum mechanics the band-width theorem appears  as the uncertainty principle,
which limits the resolution with which time and energy
(and other conjugate variables with non-commuting operators) may be observed.

\section{How time works in quantum systems}

The quantum mechanical wavefunction, $\Psi$, is used to give a
complete description of any quantum system.
If the system is dynamic, then $\Psi$ changes with time.
This is usually written as,
\begin{eqnarray}
\label{psi}
   \Psi(t_2) = U(t_2,t_1) \ \Psi(t_1) \ \ .
\end{eqnarray}
The time evolution propagator, $U(t_2,t_1)$,
describes the evolution of the system from time $t_1$ to time $t_2$.
The operator $U$ is related to the classical Green's function, $G$.
The quantum operator $U$ may be defined by \cite{MF,gw},
\begin{eqnarray}
\label{U}
U(t_2,t_1) &=& \sum_{n=0}^{\infty} \ {(-i \hbar)^n }
	\int_{t_1}^{t_2} dt^{n'} \ V(t^{n'}) \cdots
	\ \int_{t_1}^{t'''} dt'' \ V(t'')
	\int_{t_1}^{t''} dt' \   V(t')
	\nonumber \\
	&\equiv& \sum_{n=0}^{\infty} \ {(-i \hbar)^n \over n!}
	\int_{t_1}^{t_2} dt^{n'} \cdots \ \int_{t_1}^{t_2} dt''
	\int_{t_1}^{t_2} dt' \ T \ V(t^{n'}) \cdots  V(t'') V(t')
	\nonumber  \\
	&=& T e^{-i \int_{t_1}^{t_2} V(t) dt } \ \ .
\end{eqnarray}
Here $V(t)$ is the interaction that causes the system to change,
and $T$ is the Dyson time ordering operator \cite{gw}, which
imposes the causal-like constraint that $T \ V(t')V(t) = 0$ if  any $t' < t$'.
That is the time ordering operator, $T$, is proportional to
the step function, $\Theta(t_2 - t_1)$, which is 1 for $t_2 > t_1$
and 0 for $t_2 < t_1$.

Quantum time ordering corresponds to classical {\it causality}.
The invariance of physical observables under overall
reversal of time, ${\cal T}$, corresponds to classical {\it reciprocity}.
This is sometimes also called detailed balance.
This property holds in systems with no dissipation.
The time reversal operation, ${\cal T}$, corresponds to \cite{gw}
replacing $i$ by $-i$.  This leaves physical variables unchanged.
Expression of reciprocity (and possibly causality as well) is mathematically
simpler in quantum mechanics than in classical mechanics.

The time ordering operator, $T$, may be generally decomposed \cite{mg01}
into two terms, namely,
\begin{eqnarray}
\label{T}
	T = T_{av} + \Delta T \ ,
\end{eqnarray}
where $T_{av}$ is the time average value of $T$, namely $T_{av}=1$.
To understand the influence of $T_{av}$ and $\Delta T$, it is instructive
to look at their contributions in energy space.
The Fourier transform of $U(t_2,t_1)$ is the energy-space
Green's function, $G(E_0,E)$.
The step by step time-energy relation may be understood using
the Fourier transform of time ordered terms, e.g. $T V(t_2) V(t_1)$.
The Fourier transform of $\Theta(t_2 - t_1)$,
which influences the time propagation between interactions, $V(t_1)$
and $V(t_2)$, is \cite{arfken},
\begin{eqnarray}
\label{Sokhotsky}
  && \int d(t'' - t') \ \Theta(t'' -t') \ {\rm e}^{-i (E_0 - E)(t'' - t')}
	 \\ \nonumber
  && = \lim_{\eta \rightarrow 0^+} \frac{i}{E_0 - E + i \eta}
     =  \pi \delta(E_0 - E) \ -   \ \rm P_v \frac{i}{E_0 - E} \ \ .
\end{eqnarray}
The second relation is sometimes called the Sokhotsky formula.

In the formulation of scattering theory,
$\eta \to 0^+$ corresponds to the asymptotic {\it boundary}
condition\footnote{Initial conditions are usually employed in n-state
coupled channel equations.  We note that in a 2-state system
the initial condition on the amplitude $a_2(0)$ is related
to the first derivative of $a_1(0)$ since $\dot{a}_1(0) \sim V_{12}a_2(0)$.
This is a quantum example where initial conditions are used
as in classical applications of Newton's second law.}
for incoming plane waves and outgoing scattered waves \cite{gw},
and ${\rm P_v}$ is the principal value contribution that gives
contributions for $E \neq E_0$, and excludes the singular,
energy conserving point at $E = E_0$.
Since $T_{av} = 1$, the $T_{av}$ term changes nothing.
$\Delta T = T - T_{av}$ corresponds to the
time-dependent sign$(t'' - t')$ contribution to
$\Theta(t'' - t') = \frac{1}{2}(1 + {\rm sign}(t'' - t'))$,
where sign($x$) = $\pm1$ depending on whether $x$ is positive or negative.
Since $\int d(t'' - t') e^{i (E_0 - E)(t'' - t')} = 2 \pi \delta(E_0 - E)$,
it follows that,
$\frac{i}{2}\int d(t'' - t') \ {\rm sign}(t'' -t') \
{\rm e}^{-i (E_0 - E)(t'' - t')} =  {\rm P_v} \frac{1}{E_0 - E}$.
That is, the Fourier transform of $\Delta T$ is the principal value part of the 
energy propagator, which corresponds to the $i \eta$ asymptotic condition.
Thus $\Delta T$, time ordering, the direction of time propagation,
time correlation and the sequencing of interactions all correspond to 
the $i \eta$ asymptotic condition.
The direction of time propagation may be reversed by reversing 
the sign of $i \eta$, where outgoing scattered waves are replaced
by incoming scattered waves.
The absence of $i \eta$ corresponds to standing waves \cite{fanorau}.  

We emphasize that even though $\eta \to 0^+$, the influence of this term
is usually finite and can make a significant difference. 
One example is the case of the Thomas peak in electron capture \cite{mg01}
where omitting the $\eta \to 0^+$ contribution reduces the Thomas peak
by a factor of one half.  
The $i \eta$ term is like a worm in a bite from an apple:
even a very small piece has an effect.
Finite values of $i \eta$ yield dissipation with exponential decay of
probability (or exponential growth depending on the sign of $\eta$).

\begin{figure}[ht]
\label{Hanni}
\centering
%! \scalebox{0.6}{\includegraphics[viewport=75 225 800 525,clip]{mc1fig1}}
\includegraphics[height=6cm]{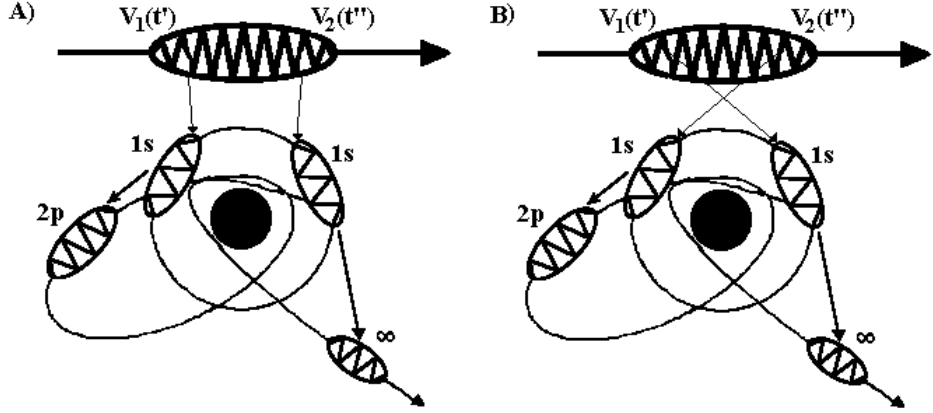}
\caption{
Time ordering between two electrons for two electrons changing 
from a (1s,1s) initial state to a ($2p,\infty$) final state.  
Time correlation occurs when $V(t'')$ is connected to $V(t')$. 
With time ordering $V(t'')V(t')$ may differ 
from $V(t')V(t'')$. 
Then $V(t'')$ is connected in time to $V(t')$.  Spatial electron
correlation occurs when two electrons interact with one another. 
}
\end{figure}

The effect of time ordering first occurs in the second order
contribution in $V$, where it may be expressed as a non-zero commutator
of $V(t)$ with $V(t')$.  
It is easily shown \cite{mg01} that,
$\Delta T \ V(t'') \ V(t') = \frac{1}{2} {\rm sign}(t'' - t') [V(t''),V(t')] $.
If the commutator $[V(t''),V(t')] = V(t'') V(t') - V(t') V(t'') = 0$, 
there is no time ordering.
Here sign$(t'' - t') = (t'' - t')/|t'' - t'|$ is like a unit vector 
that defines the direction of increasing time.
The commutator, $[V(t''),V(t')]$ provides an explicit 
time connection between interactions at different times, $t'$ and $t''$.
This non-commutivity often arises from spatial correlation \cite{mg01}.
The non-commutivity represents non-local time
entanglement between electrons, as illustrated in figure~1.  
We note that both $T_{av} V(t'') \ V(t')$
and $\Delta T  V(t'') \ V(t')$ are invariant when either
the direction of time is reversed between the two interactions,
or the overall direction of time is reversed\footnote{Of course it is the 
action of entropy (e.g. dissipation), not time ordering or causality,
that violates the invariance of physical observables under reversal of 
the flow of time.}, 
e.g. by reversing the sign of $i \eta$. 

When time ordering is present, certain phases develop\footnote{In classical
propagation \cite{MF} the principal value terms correspond to a lifetimes
decay, and are not associated with fluctuations in frequency and phase.}
in each step of the time evolution.    
Specifically, each interaction step evolves as 
$\int_{t_1}^{t_2} \ e^{i(E_I - E_{I'}) t_j}   <I|V(t_j)|I'> dt_j$ with
a phase accumulating from the $e^{i(E_I - E_{I'}) t_j}$ term.  
This modifies the phase from previous interaction steps.
Because the $e^{(E_I - E_{I'}) t_j}$ phase modulates by the
$<I|V(t_j)|I'>$ matrix element, the order of the interactions is significant.
Thus time ordering of the interaction steps leads to a specific net phase
in each order of the perturbation series.  These phases, due to
short lived quantum fluctuations in the intermediate energies
consistent with the uncertainty principle, add coherently.
If these phases are changed, e.g. by changing the sequencing of the interaction steps,
the various perturbation terms add differently.  This can affect physical 
quantities such as chemical reaction rates and cross sections.   

We note that the phases contributions from time ordering are absent 
in systems where all the unperturbed eigenenergies,%orginal%  $E_I$, are the same.
Time ordering is absent in degenerate quantum systems since in any step $\int V(t')dt'$ 
and $\int V(t'')dt''$ may be interchanged, i.e. there is no time ordering,
since $t'$ and $t''$ become interchangeable dummy variables.
Alternatively, if all the energies, $E_I$, of the basis states
are the same, then no principal value contributions 
exist since the Hilbert space of the degenerate states contains only one energy.
In a degenerate Hilbert space time ordering is suppressed. 
There are no energy fluctuations,  
no principal value contribution from $\frac{1}{E_0 - E + i \eta}$, 
no $\Delta T$ contribution, and no time ordering. 
For example if one sets $E_I = E_{av}$ in each interaction step 
(e.g. as in an eikonal or Magnus approximation), the actual time sequencing
is replaced by a time-averaged time sequence of the interactions.
On the other hand if the basis states
allow quantum energy fluctuations, i.e. the basis states are non-degenerate,
time ordering can nevertheless be zero, e.g. when $[V(t''),V(t')] = 0$.

Time ordering provides time sequencing of interactions.
We regard the removal of time ordering as rather severe:
something that can be quite important is lost.

\subsection{Experimental evidence for quantum time ordering}

\begin{figure}[ht]         % !
\centering                 % !
\includegraphics[height=8cm]{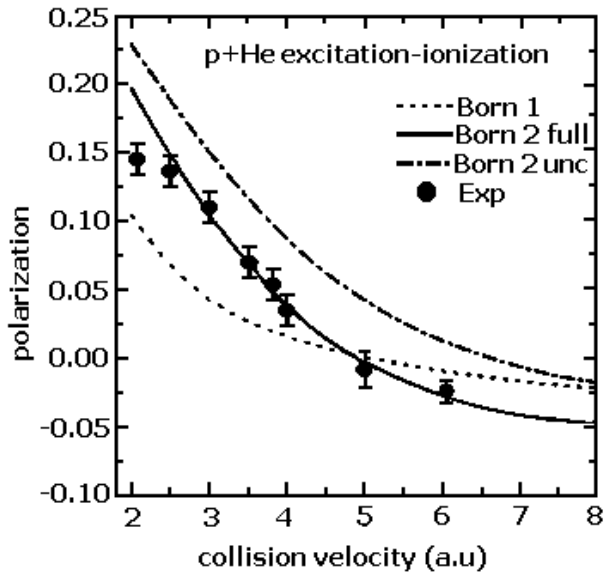}
%orginal% \centerbmp{3.9in}{3in}{c:/Jim/papers/2003/mc1fig2.bmp}
%please specify path to the figure
%orginal% \begin{figure}[ht]
\label{IEtotp}
%\centering
%\scalebox{0.5}{\includegraphics[viewport=50 200 500 650,clip]{prl01fig2}}
\caption{Calculations with and without time ordering between electrons
compared to experimental data.  Here polarized light is emitted from helium
following 1s - 2p excitation of one electron accompanied by ionization of
the second electron.  The polarization fraction is plotted as a function of
the velocity of the incident proton. The first-order calculation (Born 1)
has no time ordering.  Time ordering between electrons is omitted in our 
time-uncorrelated second-order calculation (Born 2 unc) by removing
the principal value contribution, ${\rm P_v} \frac{1}{E_0 - E}$.
The full second-order calculation (Born 2 full) includes time ordering.
}
\end{figure}

Effects due to quantum time ordering \cite{stolt93,nano02} have been 
observed in various experiments.
In one well known experiment \cite{antip} a factor of two difference in
double ionization in atomic helium was observed when incident protons
were replaced by anti-protons. Since an anti-proton may be regarded
as a proton traveling backward in time, this effect was attributed \cite{mcbook}
to a reversal of the effect of time ordering.  
Here time reversal (${\cal T}$) is compensated by charge conjugation (${\cal C}$)
so that the well accepted ${\cal CPT}$ invariance under the product of 
charge conjugation, parity change (${\cal P}$), and time reversal is preserved.  
A more direct experiment
has been done using time dependent magnetic fields to produce transitions
in Yb atoms \cite{zhao97}. 
A more recent example of time ordering was observed in the polarization 
of light emitted following
excitation with ionization of electrons in helium caused by the impact of fast
protons \cite{mgbruch03}.  Data are shown in figure~2.

\section{Discussion}

Space and time are in some ways similar. In the classical wave equation 
$x$ and $vt$ are mathematically interchangeable. 
In relativity $ict$ becomes the fourth space-time dimension.  
Many of our descriptions of classical particles
involve locating objects or events in space and time, 
as in the case of particle trajectories. 
In quantum mechanics localization is tempered by uncertainty in both space and time. 
Space and time also have some general differences. 
Space has no preferred direction.  Both causality and entropy  give time direction.  

The description of time evolution in quantum systems
relies on three principal features, namely,
i) an asymptotic condition imposed on a differential equation, 
ii) the use of dual representations, and
iii) an energy-frequency relation.  
The asymptotic condition carried by $ i \eta  \rightarrow 0^+$,
which may be imposed on the solution to the Schr\"{o}dinger equation,
gives a unique wavefunction with outgoing scattered waves.
This corresponds to a forward direction of time propagation,
imposed on the time evolution operator.
The effect of this contribution can be measured,
as we have illustrated above. 
By use of dual representations we mean that
interrelated conjugate spaces are defined by integral transforms
so that amplitudes may be analyzed using alternate representations.
For example, the Green's function $G(f',f)$ is the Fourier 
transform of the time evolution propagator $U(t_2,t_1)$.
Amplitudes related by Fourier transforms are constrained by the 
band width theorem, $\Delta f \Delta t \geq 1/4 \pi$.  
An energy-frequency relation commonly used in quantum mechanics
is linear, namely $ E = h f $, used for photons,
while the momentum-wavelength relation is an inverse relation, $p = h/\lambda$.  
In the general time dependent Schr\"{o}dinger wave equation, 
the linear energy-frequency relation corresponds to an energy 
operator\footnote{Mathematically this corresponds to 
$[E_{op},t] = \hbar$, i.e. non-commutivity.}, 
$E_{op} = i \hbar \partial / \partial t$.

Some difficulty with time ordering can arise.
In relativity, where space and time are mixed together, 
it is well known that the order in which interactions or events 
occur can be different as seen by observers who move relative to one another.  
Specifically using relativity one observer could correctly deduce that person A is 
guilty of throwing a switch that starts a fight before B throws his switch, 
while another observed moving relative to the first observed can correctly reach the 
opposite conclusion.  
In quantum mechanics the observation of events requires time
intervals greater than a minimum time interval, $\Delta t$, dictated by the uncertainty principle.
Within $\Delta t$ the sequence of events observed may not be reproducible. 
That is, causality may be violated for different reasons in relativity and in quantum 
mechanics\footnote{In relativity \cite{Jackson} events are either space-like or time-like
regions separated by the light cone at $v = c$.  
In quantum mechanics the light cone is subject to uncertainty.}.

Microscopic-macroscopic connections can arise.
When states are nearly degenerate the uncertainty principle leads to large times and distances.
For a 2s-2p transition in hydrogen, $\Delta E = 4.37 \times 10^{-6}$ eV, 
$\Delta t = 7.54 \times 10^{-11}$ sec, and $\Delta$ l = c $\Delta$ t = 2.26 cm.
Since we take $c \Delta t$ as the size of the wavepacket,
within $\Delta t$ coherence persists at macroscopically large distances
where coupling to the environment can occur.
A similar connection is provided by the asymptotic condition\footnote{Other,
less tangible examples of non-locality occur with particle identity
and EPR entanglement of quantum states.}
that $i \eta \to 0^\pm$.
Now the effect of this condition moves to
asymptotically large times\footnote{Apparently the smallest $\Delta t$ 
dominates.  That is, any envelope that damps the wavepacket
faster than other conditions will determine the minimum
size possible for an observation.  Hence this $i \eta \to 0^\pm$
condition can never set the size of $\Delta t$ in a finite universe.}.    
However, since $1/ \eta \gg \Delta t$, decoherence often occurs. 
Of course, if $i \eta$ is kept finite, then (choosing
the proper sign) the state dissipates with a lifetime of order $1/\eta$.
Decoherence and measurement are both under scrutiny at the this time.
At issue is the nature of the dynamic transition from quantum coherence
to classically well defined, observed, decoherent events.  We leave this transition
from quantum to classical physics for another discussion.

Time ordering has been recently used to define time correlation \cite{mg01},
an independent time approximation \cite{gm01} where particles evolve independently,
and conditions under which different times may be used for different particles \cite{mg03}.
This provides a framework for understanding how quantum particles and systems
of quantum particles communicate about time.
In the independent particle approximation, where spatial inter-particle interactions are
removed, use of multiple times is possible, but optional.
There is no communication between particles. 
In this limit one may use either a single time, with a single energy-time 
Fourier transform, or different times with a different 
energy-time transform for each particle.  
The use of different times for different particles is fully justified 
when coherence between single particle amplitudes is lost, 
e.g., if relatively strong randomly fluctuating residual
fields influence each particle independently.  
Then the phase coherence that is needed to synchronize quantum clocks is lost.
When spatial correlation is present, however, the use of multiple
times is not feasible, even when the evolution of the particles is uncorrelated in time \cite{mg03}.  
Thus there is an asymmetry in spatial and temporal correlation:
time correlation between particles is forbidden in the absence of spatial 
correlation, but spatial correlation between particles is permitted 
in the absence of time correlation. 

More recently we have addressed coherent electron population transfer 
by eliminating quantum time ordering \cite{sm02,msr03}.
In an ensemble of atoms with n states dynamically mixed with a strong external field, 
it can be useful to  transfer the electron populations:
where we want, when we want, for as long as we want, as often as we want,
as completely as we want, in systems as large as we want, using simple math.
This can be done in degenerate systems, i.e. systems without quantum time ordering.
If fast `kicks' are used, i.e. $V(t) \sim \delta(t-t_0)$, electron population
can be transferred from a launch state to a target state instantaneously at $t_0$
and remain there indefinitely \cite{sm02}.  Quantum transitions can occur within
the time interval, $\Delta t$, set by the uncertainty relation,
but no observation of this transition can be reproducibly made within $\Delta t$. 
This has application in coherent population trapping of electrons
useful in quantum computing, in electromagnetically induced transparency,
may be useful in slowing and stopping of light, and in genetic learning algorthims
used to control chemical reactions.
More details are given elsewhere in this book \cite{rsm03}. 

\begin{figure}[ht]
\label{Action}
\centering
\includegraphics[height=6cm]{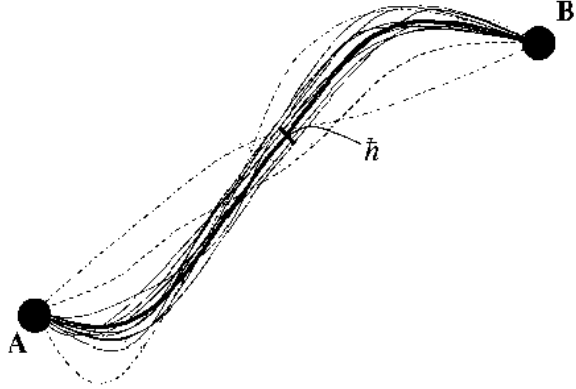}
\caption{The physical path from A to B is determined by the extreme
value of the action integral (or phase area), $A = \int V(t) dt$.
In a classical system this path is unique.
In a quantum system the possible pathways are broadened by
the uncertainty principle.  This corresponds to a
quantum freedom of uncertainty.  
If the action integral is an integer value of $\hbar/2$,
the revivals may occur periodically.}
\end{figure}

In classical physics the trajectory of a particle is constrained by 
Fermat's principle:  that nature seeks
the most (or possibly the least) efficient way to go from A to B.  
The particle's unique path is determined by minimizing the action.
This is illustrated by the dark line in figure~3.
On the other hand quantum mechanics may be obtained from classical 
mechanics \cite{Messiah} by quantizing the action.
In quantum mechanics many paths from A to B may be possible, although 
those outside of an envelope of trajectories,  whose width is 
proportional to $\hbar$, are statistically improbable.
The $\Delta E$ in energy means that there are quantum fluctuations in energy,
which briefly violate conservation of energy.  
This leads to time correlations \cite{gm02} insuring that the electrons cooperate 
in seeking the most efficient way to get from A to B, subject to both the
constraint of classical least action and the freedom of quantum uncertainty.  

We note that our approach to characterizing how time works
in quantum systems differs somewhat from Briggs and Rost \cite{br00}
who emphasize the widespread use of a classical $V(t)$ in
the time dependent Schr\"{o}dinger equation to emphasize the similarity
of time in classical and quantum systems.  We emphasize the differences.

\section{Summary}

Time works somewhat differently in quantum mechanics than in classical mechanics.
Unlike in classical mechanics time is not a well defined physical observable in quantum mechanics.
Also in quantum mechanics there are fluctuations in phase occuring in each interaction step  
that enforce time ordering.
While time itself is presumably the same quantum mechanically and classically,
in quantum mechanics time operates in different ways.

\section{Acknowledgments}
We thank A.R.P. Rau and C. Rangan for valuable comments.
KhR is supported by a NSF-NATO Fellowship.


\begin{thebibliography}{99}
\bibitem{MF} P.M. Morse and H. Feshbach, {\it Methods of Theoretical Physics.} McGraw-Hill, NY, 1953.
\bibitem{Jackson} J.D. Jackson, {\it Classical Electrodynamics}. John Wiley, NY, 1975.
\bibitem{mair02} A. Mair, J. Hager, D.F. Phillips, R.L. Wadsworth, and M.D. Lukin, {\it Phys. Rev A} {\bf 65}, 031802 (2002).
\bibitem{gw} M.L. Goldberger and K.M. Watson, {\it Collision Theory}. John Wiley, NY, 1964.
\bibitem{mg01} J.H. McGuire, A.L. Godunov, S.G. Tolmanov, Kh.Kh. Shakov, H. Schmidt-Bocking, and R.M. Dreizler, {\it Phys. Rev. A} {\bf 63}, 052706 (2001).
\bibitem{arfken} G.B. Arfken and H.J. Weber, {\it Mathematical Methods for Physicists}. Academic Press, NY, 1995.
\bibitem{fanorau} U. Fano and A.R.P. Rau, {\it Atomic Collisions, Spectra}. Academic Press, Orlando, 1986.
\bibitem{stolt93} N. Stoltfoht, {\it Phys. Rev. A} {\bf 48}, 2980 (1993).
\bibitem{nano02} J.H. McGuire and A.L. Godunov, {\it Proceedings of the International Meeting on Electron Scattering from Atoms, Nuclei, Molecules, Bulk Matter}. Eds. C. Whelan and N. Mason. Kluwer Academc Press, NY, 2002.
\bibitem{antip} L.H. Anderson, P. Hvelplund, H. Knudsen, H.P. Moller, K. Elsner, K.G. Rensfelt, and E. Uggerhoj, {\i Phys. Rev. Lett.} {\bf 57}, 2147 (1986).
\bibitem{mcbook} J.H. McGuire, {\it Electron Correlation Dynamics in Atomic Collisions}. University Press, Cambridge, 1997.
\bibitem{zhao97} H.Z. Zhao, Z.H. Lu, and J. E. Thomas, {\it Phys. Rev. Lett.} {\bf 79}, 613 (1997).
\bibitem{mgbruch03} J.H. McGuire, A.L. Godunov, Kh.Kh. Shakov, H. Merabet, J. Hanni, and R. Bruch, {\it J. Phys. B} {\bf 36}, 209 (2003).
\bibitem{gm01} A.L. Godunov and J.H. McGuire, {\it J. Phys. B} {\bf 34}, L223 (2001). 
\bibitem{mg03} J.H. McGuire and A.L. Godunov, {\it Phys. Rev. A} {\ bf 67}, 041703 (2003).
\bibitem{sm02} Kh.Kh. Shakov and J.H. McGuire, {\it Phys. Rev. A} {\ bf 67}, 033405 (2003).
\bibitem{msr03} J.H. McGuire, Kh.Kh. Shakov, and Kh. Yu. Rakhimov, {\it J. Phys. B} {\bf 36}, 3145 (2003).
\bibitem{rsm03} Kh.Yu. Rakhimov, Kh.Kh. Shakov, and J.H. McGuire, elsewhere in this volume, 2003.
\bibitem{Messiah} A. Messiah, {\it Quantum Mechanics}. North-Hollland Publishing Co., Amsterdam, 1961.
\bibitem{gm02} A.L. Godunov, J.H. McGuire, Kh.Kh. Shakov, H. Merabet, J. Hanni, R. Bruch, and V. Schipakov, {\it J. Phys. B} {\bf 34}, 5055 (2001).
\bibitem{br00} J.S. Briggs and J.M. Rost, {\it Euro. Phys. J. D} {\bf 10}, 311 (2000).
\end{thebibliography}
\end{document}